\documentstyle[12pt]{article}
\begin{document}
\title{Critical Behavior of the Widom-Rowlinson Lattice Model}
\author{Ronald Dickman$^*$\\
        Department of Physics and Astronomy\\
        Lehman College, CUNY\\
        Bronx, NY 10468-1589\\
        and\\
        George Stell$^{\dag}$\\
        Department of Chemistry\\
        State University of New York at Stony Brook\\
        Stony Brook, NY 11794-3400\\}

\maketitle
\begin{abstract}
We report extensive Monte Carlo simulations of the Widom-Rowlinson
lattice model in two and three dimensions.  Our results yield precise
values for the critical activities and densities, and clearly place the
critical behavior in the Ising universality class.
\end{abstract}

$^*$ e-mail address: dickman@lcvax.lehman.cuny.edu\\
$^{\dag}$ e-mail address: gstell@sbchm1.chem.sunysb.edu

\newpage

The Widom-Rowlinson (WR) hard-sphere mixture
is perhaps the simplest binary fluid model
showing a continuous unmixing transition, and has been the subject of
considerable study regarding its thermal and  interfacial properties
\cite{wr70,wrbk}, as has the Gaussian $f$-function version of the model
introduced somewhat earlier by Helfand and Stillinger \cite{hlst}.
Despite this interest, however, definitive results on the location and nature
of the WR critical point are lacking.  Indeed, except for the
essentially inconclusive series analyses of the Gaussian version
\cite{hlst,rwbar},
little has appeared by way of precise quantitative analysis on the critical
behavior
of any WR-type model.  As a first step in this direction we have performed
extensive simulations of the lattice-gas analog of the WR hard-sphere
mixture the
Widom-Rowlinson lattice model (WRL) \cite{wrbk}.

In the original WR model, AB pairs interact {\em via} a hard-sphere
potential whilst AA and BB pairs are noninteracting.
By the WRL model we mean a two-component lattice gas in
which sites may be at most singly occupied, and in which nearest-neighbor A-B
pairs are forbidden.  Like the WR model, this is evidently an athermal model
(all allowed configurations
are of the same energy), and is characterized solely by the densities of the
two species or the corresponding chemical potentials
$\mu_A$ and $\mu_B$.  ($\mu_A = \mu_B = \mu_c$ of course, at the critical
point.)
The WRL may be viewed as an extreme member of a family of {\em binary
alloy models}.  The Ising model, as is known, may be transcribed into such a
model by identifying up and down spins with A and B particles, resp., yielding
a ``close-packed" alloy which unmixes at the Ising critical temperature.
Allowing a small fraction of vacant sites results in a ``dilute
binary alloy" (DBA) with a
somewhat depressed critical temperature; continuing the dilution process, one
arrives at a model with $T_c = 0$.  This zero-temperature terminus of the DBA
critical line is precisely the WRL critical point.  One is then led to
ask whether
the entire line shares a common critical behavior, or whether its character
changes at some point.  Although the former expectation is clearly favored
on the basis of universality, a careful examination of this question
nevertheless
appears worthwhile.  One should also note that whilst in the WRL there is no
temperature {\em per se}, $\mu_A + \mu_B$  is a
temperature-like variable, so that along the symmetry line $\mu_A =
\mu_B = \mu$,
the susceptibility scales as $\chi \simeq (\mu - \mu_c)^{-\gamma} $, the order
parameter, $\rho_A - \rho_B \simeq (\mu - \mu_c)^{\beta} $, $(\mu > \mu_c)$,
and so on \cite{hlst}.  ($h \equiv \mu_A - \mu_B $ plays the role of an
external field.)

Our simulations employ periodic square or cubic lattices of side $L$ ($L$ =
10, 20,
40, 80, and 160 in two dimensions; 8, 12, 16, 24, 40, and 64 in three), and are
performed in the grand canonical ensemble, along the symmetry line, $z_A =
z_B = z \equiv e^{\mu /kT}$.  The state of site i is conveniently represented
by $\sigma_{\rm i} = 1, 0$, or -1, corresponding to i occupied by A,
vacant, or occupied by B.  Three kinds of moves are employed:

\noindent (i) A ``flip" in which,
if $\sigma_{\rm i} \neq 0$, $\sigma_{\rm i} \rightarrow -\sigma_{\rm i}$
or 0 with equal likelihood, while if $\sigma_{\rm i} = 0$, $\sigma_{\rm i}
\rightarrow
\pm 1$ with equal likelihood.
Moves
vacating a site are accepted with probability $\min[1,1/z]$, those adding a new
particle with probability $\min[1,z]$, provided no A-B nearest neighbor pairs
are
formed.  Acceptance of changes in species is solely contingent on the latter
condition.

\noindent (ii) Exchange of the states of a pair of randomly chosen
sites (anywhere
in the system).  Acceptance again depends on no A-B pair being generated.

\noindent (iii) A ``cluster flip" in which the entire particle cluster
connected to site
i is changed to the opposite species.  Since clusters are bounded exclusively
by vacant sites, such moves never generate A-B pairs and are always accepted.

In processes (i) and (iii) site i is chosen at random; of course if
$\sigma_{\rm i} = 0$
there is no cluster flip.  Cluster flips are of particular
simplicity in WR models,
and allow for rapid relaxation of any species imbalance.  They are, however,
time-consuming,
particularly in large systems near or above the critical point,
since an appreciable
fraction of sites may belong to a single cluster.  We found a
reasonably efficient
procedure was to attempt such moves about ten times per lattice update (lud).
(A lud denotes $L^d$ moves of types (i) and (ii).)  Block averages
(over 100 lud's) of mole fractions were then equal to within about in
part in $10^3$.
Our simulations furnish the following quantities as a function of $z$:
the density
$\rho$ and its variance; the order parameter $m \equiv |\rho_a - \rho_b|$
and its variance; the probability distribution $P(\Delta)$  of the population
excess $\Delta \equiv N_A - N_B$;
the reduced fourth cumulant, $U(z,L) \equiv 1 -<m^4>/3<m^2>$ \cite{bin81}.

We used two independent methods to locate the critical activity $z_c$:
order-parameter histograms and fourth-cumulant crossings.  The former method
is based on the observation that $P(\Delta)$  undergoes a clear change in
character at a certain activity $z_c(L)$: for $z < z_c(L)$, $P(\Delta)$ is
peaked at zero, whilst above $z_c(L)$ it is bimodal; just at
$z_c$ it has a broad plateau.  Finite-size scaling ideas \cite{fishfss,privfss}
imply that as $L\ \rightarrow \infty$,
\begin{equation}
z_c(L) \approx z_c + AL^{-1/\nu}.
\label{zcl}
\end{equation}
In analyzing the data we adopt, as a working hypothesis, the
Ising class $\nu$-value.  Plotting the two-dimensional $z_c(L)$ data
{\em vs.} $L^{-1}$ yields a straight line with intercept $z_c = 2.0644$,
and a similar analysis in 3-d (using $\nu = 0.629$ \cite{ferrland})
gives $z_c = 0.7842$.
The widely-employed fourth-cumulant analysis uses the fact that
$U(z_c,L)$ rapidly converges to a nontrivial value $U^*$ with
increasing $L$ \cite{bin81}.  In our studies the crossings of the three
largest $L$-values agree to within uncertainty.  In 2d the crossing is at
$z_c = 2.0636$, $U^* = 0.607$, while in 3d we find $z_c = 0.7840$,
$U^* = 0.478(5)$.  Our final estimates for the critical activities are:
\begin{equation}
z_c = 2.0604(4) \;\;\;\;\;\;\;\; (2d),
\end{equation}
\begin{equation}
z_c = 0.7841(1) \;\;\;\;\;\;\;\; (3d).
\end{equation}
The corresponding critical densities are
$\rho_c = 0.618(1) $ (2d) and 0.3543(1) (3d).
(The Bethe-Guggenheim approximation yields 0.429 and
0.2941, respectively, for the critical densities in 2 and 3d \cite{wrbk})
The hypothesis of Ising-like critical behavior
finds significant support in the fact that the $z_c(L)$ plots are linear
(this confirms the assumed value of $\nu$), and from the
good agreement of our $U^*$ with the accepted values of 0.61 in 2d
\cite{bin81} and 0.47 in 3d \cite{ferrland}.
Further support appears when we examine the order
parameter and susceptibility.

In Figs. 1 and 2 we plot, for 2d and 3d, respectively,
the scaled order parameter, $\tilde{m} = L^{\beta/\nu} m$
{\em vs.} the scaling variable $t = L^{-1/\nu} |z-z_c|$, using
the Ising model values for the exponents.  The data are seen
to collapse nicely, and for 2d are fully consistent with $\beta = 1/8$,
i.e., the slope of the scaling plot is $\approx 1/8$ for $z>z_c$,
and $\approx -7/8$ for $z < z_c$.
The slope of the 3d data is a bit low, a fit to the most linear
portion yielding about 0.31, as compared with the accepted value
$\beta = 0.326(4)$ \cite{ferrland}.  Finite size effects seem the most likely
cause for this small discrepancy.
The susceptibility is given by $\chi \equiv \partial <\Delta>/\partial h =
\frac{1}{2} var[\Delta] $.  To study its scaling we plot $\tilde{\chi} =
L^{-\gamma/\nu} $ {\em vs.} $t = L^{-1/\nu} |z - z_m(L)|$, where $z_m(L)$ is
the activity at which $\chi (z,L)$ takes its maximum.  In this case
there is a reasonably good data collapse and the slopes of the
scaling plots are in good agreement with the accepted values of $\gamma$,
as seen in Figs. 3 and 4.
In WR systems the analog of the specific heat is the
compressibility, given by $ \kappa = L^d var[\rho]/\rho^2$.  We are unable
to obtain $\kappa$ to good precision in the critical region (similar difficulty
is encountered for the specific heat in Ising simulations), but can at least
report that in two dimensions the maximum
compressibility grows $\propto \ln L$, as expected for models in the Ising
class.  Finally, it is of some interest to know the shape of the coexistence
curves in two and three dimensions; these are shown in Fig. 5.

In summary, we have performed extensive and detailed simulations of the
Widom-Rowlinson lattice model in two and three dimensions, and find strong
evidence of Ising-like critical behavior.
We believe that further studies of other aspects of the WRL are in order, for
example, surface free energies and roughening, critical dynamics,
and the influence of a driving field.  Series analysis of critical behavior
also appears feasible and worthwhile.
Such investigations promise to
shed new light on the Ising universality class, as well as on critical behavior
in binary fluids.

R.D. acknowledges the support of the Division of Chemical Sciences,
Offices of Basic Energy Sciences, Office of Energy Research, U.S.
Department of Energy, and G.S. acknowledges the support of the
National Science Foundation.

\newpage

\noindent{\bf Figure Captions}

\vspace{1em}
\noindent Fig. 1. Scaling plot of the order parameter, 2d.
Open circles: $L=20$;
open diamonds: $L=40$; filled circles: $L=80$; squares: $L=160$.
The straight lines
have slopes of 1/8 and -7/8.
\vspace{1em}

\noindent Fig. 2.  Scaling plot of the order parameter, 3d.
Open diamonds: $L=8$;
open squares: $L=12$; open circles: $L=16$; filled circles:
$L=24$; filled squares:
$L=40$.  The line has slope 0.326.
\vspace{1em}

\noindent Fig. 3.  Scaling plot of the susceptibility, 2d.
Symbols as in Fig. 1.
The slope of the line is -1.75.
\vspace{1em}

\noindent Fig. 4. Scaling plot of the susceptibility, 3d.
Symbols as in Fig. 2.
The slope of the line is -1.247.
\vspace{1em}

\noindent Fig. 5. Coexistence curves in 2d (broken line) and
3d (solid line).
\vspace{1em}

\end{document}